%% file: bg19.tex
\documentclass[aps,prl,twocolumn,showpacs,floatfix,longbibliography]{revtex4-1}
\usepackage{graphicx,amsfonts,amssymb,amsmath}
\usepackage{textcomp}
\usepackage{esint}
\usepackage[colorlinks,linktocpage,bookmarks=false,citecolor=blue,linkcolor=red,urlcolor=blue]{hyperref}
\usepackage[T1]{fontenc}
\usepackage[dvipsnames]{xcolor}
\usepackage[english]{babel}
\usepackage{afterpage}

\allowdisplaybreaks

\begin{document}
	
	\graphicspath{{./figures_submit/}}
	\title{Glassy properties of the Bose-glass phase of a one-dimensional disordered Bose fluid}
	
	\author{Nicolas Dupuis}
	\affiliation{Sorbonne Universit\'e, CNRS, Laboratoire de Physique Th\'eorique de la Mati\`ere Condens\'ee, LPTMC, F-75005 Paris, France}
	
	\date{September 16, 2019} 
	
	\begin{abstract}
		We study a one-dimensional disordered Bose fluid using bosonization, the replica method and a nonperturbative functional renormalization-group approach. The Bose-glass phase is described by a fully attractive strong-disorder fixed point characterized by a singular disorder correlator whose functional dependence assumes a cuspy form that is related to the existence of metastable states. At nonzero momentum scale, quantum tunneling between these metastable states leads to a rounding of the nonanalyticity in a quantum boundary layer that encodes the existence of rare superfluid regions responsible for the $\omega^2$ behavior of the (dissipative) conductivity in the low-frequency limit. These results can be understood within the ``droplet'' picture put forward for the description of glassy (classical) systems.
	\end{abstract}
	
	\maketitle
	
	\input{./definition.tex}
	
	\paragraph{Introduction.} In quantum many-body systems the competition between interactions and disorder may lead to a localization transition. In an electron system disorder can turn a metal into an Anderson insulator~\cite{Anderson_50years} (or an electron glass~\cite{Davies82,Mueller04,Somoza08,Vaknin00,Amir11} in the presence of long-range Coulomb interactions) as a result of the single-particle wave-function localization. In a boson system the transition occurs between a superfluid phase and a localized phase~\cite{Giamarchi87,*Giamarchi88} dubbed Bose glass~\cite{Fisher89}. These disorder-dominated phases are all characterized by the absence of transport, i.e. a vanishing conductivity in the limit of zero frequency. The insulating behavior is however only one of the fundamental properties of these localized phases. As in disordered classical systems, one also expects ``glassy'' properties due to the existence of metastable states. The understanding of these glassy properties is a key issue in the physics of disordered quantum many-body systems~\cite{Lemarie19}.
	
	Most of our understanding of disordered (glassy) classical systems comes from the replica approach~\cite{Edwards75} and Parisi's ``replica-symmetry-breaking'' scheme~\cite{Mezard87} or the functional renormalization group (FRG)~\cite{Fisher85,Narayan92,Nattermann92,Chauve01,Ledoussal04,Tarjus04}. In the latter approach a crucial feature is that the disorder correlator assumes a cuspy functional form whose origin lies in the existence of many different microscopic, locally stable, configurations~\cite{Balents96}. This metastability leads in turn to a host of effects specific to disordered systems: non-ergodicity, pinning and ``shocks'' (static avalanches), depinning transition and avalanches, chaotic behavior, slow dynamics and aging, etc. In this paper we show that the (nonperturbative version of the) FRG approach gives a fairly complete description of the  Bose-glass phase of a one-dimensional disordered Bose fluid in agreement with the phenomenological ``droplet'' picture put forward for glassy (classical) systems~\cite{Fisher88b}.  
	
\begin{figure}
		\centerline{\includegraphics[width=7.5cm,angle=0]{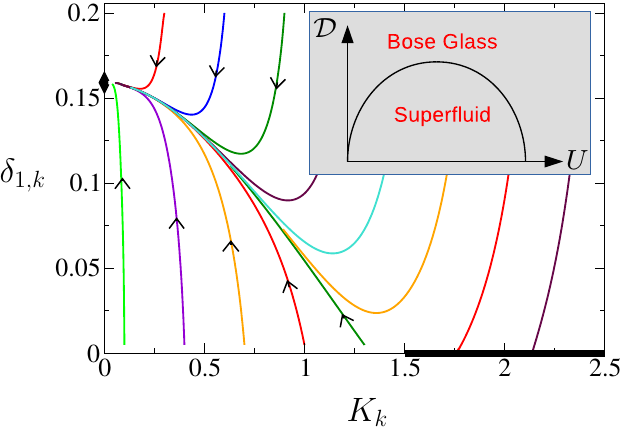}}
		\caption{Flow diagram projected onto the plane $(K_k,\delta_{1,k})$ where $K_k$ is the running Luttinger parameter and $\delta_{1,k}$ is the first harmonic of the dimensionless potential $\delta_k(u)=-(K^2/v^2k^3)V_k''(u)=\sum_{n=1}^\infty \delta_{n,k}\cos(2nu)$. The thick solid line $(K_k\geq 3/2,\delta_{1,k}=0)$ shows the attractive line of fixed points corresponding to the superfluid phase and the black dot $(K^*=0,\delta^*_1\simeq 0.159)$ the attractive fixed point corresponding to the Bose-glass phase. The inset shows the schematic phase diagram of a disordered one-dimensional Bose fluid as a function of the boson repulsion $U$ and the disorder $\cal D$.}
		\label{fig_flow_diagram} 
	\end{figure}

	The competition between disorder and interactions in one-dimensional disordered boson systems was first addressed by Giamarchi and Schulz by means of a perturbative RG approach~\cite{Giamarchi87,*Giamarchi88,Ristivojevic12,*Ristivojevic14}. They showed that whenever the (dimensionless) Luttinger parameter $K$, which characterizes the quantum fluctuations of the particle density ($K\to 0$ corresponds to the classical limit), is smaller than 3/2 even an infinitesimal disorder results in localization and thus destroys the superfluid phase (see the inset of Fig.~\ref{fig_flow_diagram} for the generic phase diagram of a one-dimensional disordered Bose fluid). Scaling arguments have led to the conclusion that the Bose-glass phase also exists in higher dimension and is generically characterized by a nonzero compressibility, the absence of a gap in the excitation spectrum and an infinite superfluid susceptibility~\cite{Fisher89}. Experimentally, the superfluid--Bose-glass transition has regained a considerable interest owing to the observation of a localization transition in cold atomic gases~\cite{Billy08,Roati08,Pasienski10} as well as in magnetic insulators~\cite{Hong10,Yamada11,Zheludev13}. The Bose-glass phase is also relevant for the physics of one-dimensional Fermi fluids~\cite{Giamarchi_book}, 
	charge-density waves in metals~\cite{Fukuyama78} and superinductors~\cite{Houzet19}. 
	
	Whereas the critical behavior at the superfluid--Bose-glass transition, which is of Berezinskii-Kosterlitz-Thouless type, is well understood in the weak-disorder limit~\cite{[{In the limit of strong disorder, the superfluid--Bose-glass transition is still thought to be of Berezinskii-Kosterlitz-Thouless type but with a critical Luttinger parameter $K_c$ that differs from 3/2. For a recent discussion, see }]Doggen17}, the perturbative RG does not allow one to study the localized phase where disorder flows to strong coupling. Using bosonization, the replica method and a nonperturbative functional renormalization-group approach, we find that the Bose-glass phase is described by a fully attractive strong-disorder fixed point characterized by a vanishing Luttinger parameter $K^*=0$ and a singular disorder correlator which assumes a cuspy functional form. At nonzero momentum scale $k$, as a consequence of quantum fluctuations, the cusp singularity is rounded in a quantum boundary layer whose size depends on an effective Luttinger parameter $K_k\sim k^\theta$ which vanishes with an exponent $\theta=1/2$, thus yielding a dynamical critical exponent $z=3/2$. Many of these results are similar to those obtained within the FRG approach to classical systems where temperature plays the role of the Luttinger parameter (the RG flow is attracted by a zero-temperature fixed point), although usually in higher space dimensions. This reveals some of the glassy properties of the Bose-glass phase: metastability, pinning and ``shocks'' (or static avalanches) but also emphasizes the crucial role of quantum tunneling between different metastable configurations. The latter leads to the existence of rare superfluid regions that are responsible for a (dissipative) conductivity vanishing as $\w^2$ in the low-frequency limit.  
	
	\paragraph{FRG approach.} 
	We consider one-dimensional interacting bosons with Hamiltonian $\hat H_0+\hat H_{\rm dis}$. In the absence of disorder, at low energies the system is described by the Tomonaga-Luttinger Hamiltonian~\cite{Giamarchi_book,Haldane81,Cazalilla11}
	\begin{equation}
	\hat H_0 = \int dx \frac{v}{2\pi} \left\{ \frac{1}{K} (\dx \hat\varphi)^2 + K (\dx \hat\theta)^2 \right\}  
	\end{equation}
	(we set $\hbar=k_B=1$),
	where $\hat\theta$ is the phase of the boson operator $\hat\psi(x)=e^{i\hat\theta(x)}\hat\rho(x)^{1/2}$ and $\hat\varphi$ is related to the density {\it via} $\hat\rho=\rho_0 - \frac{1}{\pi}\dx\hat\varphi+2\rho_2 \cos(2\pi\rho_0 x-2\hat\varphi)$ ($\rho_0$ is the average density). $\hat\varphi$ and $\hat\theta$ satisfy the commutation relations $[\hat\theta(x),\partial_y\hat\varphi(y)]=i\pi\delta(x-y)$. $v$ denotes the sound-mode velocity and the dimensionless quantity $K$, which encodes the strength of boson-boson interactions, is the Luttinger parameter. The disorder contributes to the Hamiltonian a term $\hat H_{\rm dis}=\int dx V(x) \hat\rho(x)$ where the random potential $V(x)$ is assumed to have a Gaussian probability distribution with zero mean and variance $\overline{V(x)V(x')}=(\calD/\rho_2^2)\delta(x-x')$ (an overline indicates disorder averaging). The average over disorder can be done using the replica method, i.e. by considering $n$ copies of the model. This leads to the following Euclidean action (after integrating out the field $\theta$)~\cite{Giamarchi87,*Giamarchi88}  
	\begin{multline}
	S = \sum_a \int dx \inttau \frac{v}{2\pi K} \left\{ (\dx\varphi_a)^2 + \frac{(\dtau\varphi_a)^2}{v^2} \right\} \\ 
	-\calD \sum_{a,b} \int dx \inttau \, d\tau' \cos[2\varphi_a(x,\tau)-2\varphi_b(x,\tau')] , 
	\label{action} 
	\end{multline}
	where $\varphi_a(x,\tau)$ is a bosonic field with $\tau\in[0,\beta]$ an imaginary time ($\beta=1/T\to\infty$) and $a,b=1\cdots n$ are replica indices. If we interpret $y=v\tau$ as a space coordinate, the action~(\ref{action}) also describes (two-dimensional) elastic manifolds in a (three-dimensional) disordered medium~\cite{Fisher86,Chauve00,Ledoussal02a,Ledoussal04,Ledoussal06a,Balents04}, yet with a periodic structure and a perfectly correlated disorder in the $y$ direction~\cite{Balents93,Giamarchi96,Fedorenko08}. The Luttinger parameter, which controls quantum fluctuations in the Bose fluid, defines the temperature of the classical model.  
	
	Most physical quantities can be obtained from the free energy $-\ln\calZ[J]$ (the logarithm of the partition function) or, equivalently, from the effective action (or Gibbs free energy) 
	\begin{equation}
	\Gamma[\phi] = - \ln\calZ[J] + \sum_a \int dx \inttau J_a \phi_a 
	\end{equation}
	defined as the Legendre transform of $\ln\calZ[J]$. Here $J_a$ is an external source which couples linearly to the $\varphi_a$ field and allows us to obtain the expectation value $\phi_a(x,\tau)=\mean{\varphi_a(x,\tau)}=\delta\ln\calZ[J]/\delta J_a(x,\tau)$. We compute $\Gamma[\phi]$ using a Wilsonian nonperturbative FRG approach~\cite{Berges02,Delamotte12} where fluctuation modes are progressively integrated out~\cite{not33}.
	This defines a scale-dependent effective action $\Gamma_k[\phi]$ which incorporates fluctuations with momenta (and frequencies) between a running momentum scale $k$ and a UV scale $\Lamb$. The effective action of the original model, $\Gamma_{k=0}[\phi]$, is obtained when all fluctuations have been integrated out whereas $\Gamma_\Lamb[\phi]=S[\phi]$. $\Gamma_k$ satisfies a flow equation which allows one to obtain $\Gamma_{k=0}$ from $\Gamma_\Lamb$ but which cannot be solved exactly. 
	
	A possible approximation scheme is to expand the effective action
	\begin{equation}
	\Gamma_k[\phi] = \sum_a \Gamma_{1,k}[\phi_a] - \half \sum_{a,b} \Gamma_{2,k}[\phi_a,\phi_b] + \cdots 
	\end{equation}
	in increasing number of free replica sums and to truncate the expansion to a given order~\cite{Tarjus08,Tissier08}. In the following we retain only $\Gamma_{1,k}$ and $\Gamma_{2,k}$ and consider the ansatz 
	\begin{equation}
	\begin{split} 
	&\Gamma_{1,k}[\phi_a] = \int dx \inttau \llbrace \frac{Z_{x}}{2} (\dx\phi_a)^2 + \half\phi_a \Delta_k(-\dtau) \phi_a \rrbrace , \\ 
	&\Gamma_{2,k}[\phi_a,\phi_b] = \int dx \inttau\, d\tau'\, V_k(\phi_a(x,\tau)-\phi_b(x,\tau')) , 
	\end{split}
	\label{ansatz}
	\end{equation}
	with initial conditions $\Delta_\Lamb(i\w)=\w^2/\pi vK$ and $V_\Lamb(u)=2\calD \cos(2u)$. Here $\w\equiv\wn=2\pi nT$ ($n$ integer) is a Matsubara frequency (we drop the index $n$ since $\wn$ becomes a continuous variable in the limit $T\to 0$). The $\pi$-periodic function $V_k(u)$ can be interpreted as a renormalized second cumulant of the disorder. The statistical tilt symmetry (STS)~\cite{Schulz88,not55} implies that $Z_{x}=v/\pi K$ remains equal to its initial value and no higher-order space derivatives are allowed. As for the part involving time derivatives, we assume a quadratic form with an unknown ``self-energy'' $\Delta_k(i\w)$ satisfying $\Delta_k(i\w=0)=0$ as required by the STS~\cite{[{This quadratic approximation involving time derivatives to inifinite order is known as the LPA$''$, see }]Hasselmann12,*Rose18}. 
	By inserting the ansatz~(\ref{ansatz}) into the (exact) flow equation satisfied by $\Gamma_k[\phi]$ we obtain coupled RG equations for $\Delta_k(i\w)$ and $V_k(u)$. We refer to the Supplemental Material for more detail about the implementation of the FRG approach and the explicit expression of the flow equations~\cite{not11}. 
	
	In the weak-disorder limit it is sufficient to approximate $\Delta_k(i\w)=Z_x\w^2/v_k^2$ and $V_k(u)=2\calD_k \cos(2u)$. The flow equations for the velocity $v_k$, the Luttinger parameter $K_k=v_k/\pi Z_x$ and $\calD_k$ encompass the one-loop equations derived by Giamarchi and Schulz~\cite{Giamarchi87,*Giamarchi88}. One finds an attractive line of fixed points for $\calD=0$ and $K>3/2$ corresponding to the superfluid phase where the system is a Luttinger liquid. The line $\calD=0$ becomes repulsive when $K<3/2$; $\calD_k$ then flows to strong coupling which signals the Bose-glass phase. The transition between the superfluid and Bose-glass phases is in the Berezinskii-Kosterlitz-Thouless universality class. 
	
	\paragraph{Bose-glass phase.}
	The nonperturbative FRG approach allows us to follow the flow into the strong-disorder regime and determine the physical properties of the Bose-glass phase. All trajectories that do not end up in the superfluid phase are attracted by a fixed point characterized by a vanishing Luttinger parameter $K^*=0$ and a singular potential that exhibits a cusp at $u=n\pi$ ($n$ integer) in its second derivative (written here in a dimensionless form),  
	\begin{equation}
	\delta^*(u) = - \frac{K^2}{v^2} \lim_{k\to 0} \frac{V_k''(u)}{k^3} 
	= \frac{1}{2 a_2} \left[ \left( u - \frac{\pi}{2} \right)^2 - \frac{\pi^2}{12} \right] 
	\label{deltaFP}
	\end{equation}
	for $u\in [0,\pi]$, where $a_2$ is a nonuniversal number. The flow diagram obtained from the numerical solution of the flow equation and projected onto the plane $(K_k,\delta_{1,k})$, where $\delta_{1,k}$ is the first harmonic of $\delta_k(u)=\sum_{n=1}^\infty \delta_{n,k}\cos(2nu)$, is shown in Fig.~\ref{fig_flow_diagram}.
	
	The vanishing of $K_k\sim k^\theta$ is controlled by an exponent $\theta=z-1$ which is related to the dynamical critical exponent $z$ at the Bose-glass fixed point. It is difficult to predict precisely the values of $z$ and $\theta$, which turn out to be sensitive to the RG procedure, but we will argue below that $z=3/2$ and $\theta=1/2$. The vanishing of the Luttinger parameter has important consequences. First, it implies that the charge stiffness (or Drude weight) $D_k=v_kK_k=(v/K)K_k^2\sim k^{2\theta}$, i.e. the weight of the zero-frequency delta peak in the conductivity, vanishes for $k\to 0$ in the Bose-glass phase whereas the compressibility $\kappa=1/\pi^2Z_x=K/\pi v$ is unaffected by disorder. Second, it shows that quantum fluctuations are suppressed at low energies. We thus expect the phase field $\varphi(x,\tau)$ to have weak temporal (quantum) fluctuations and to adjust its value in space so as to minimize the energy due to the random potential, a hallmark of pinning.
	
	\begin{figure}
		\centerline{\includegraphics[width=3.7cm,angle=0]{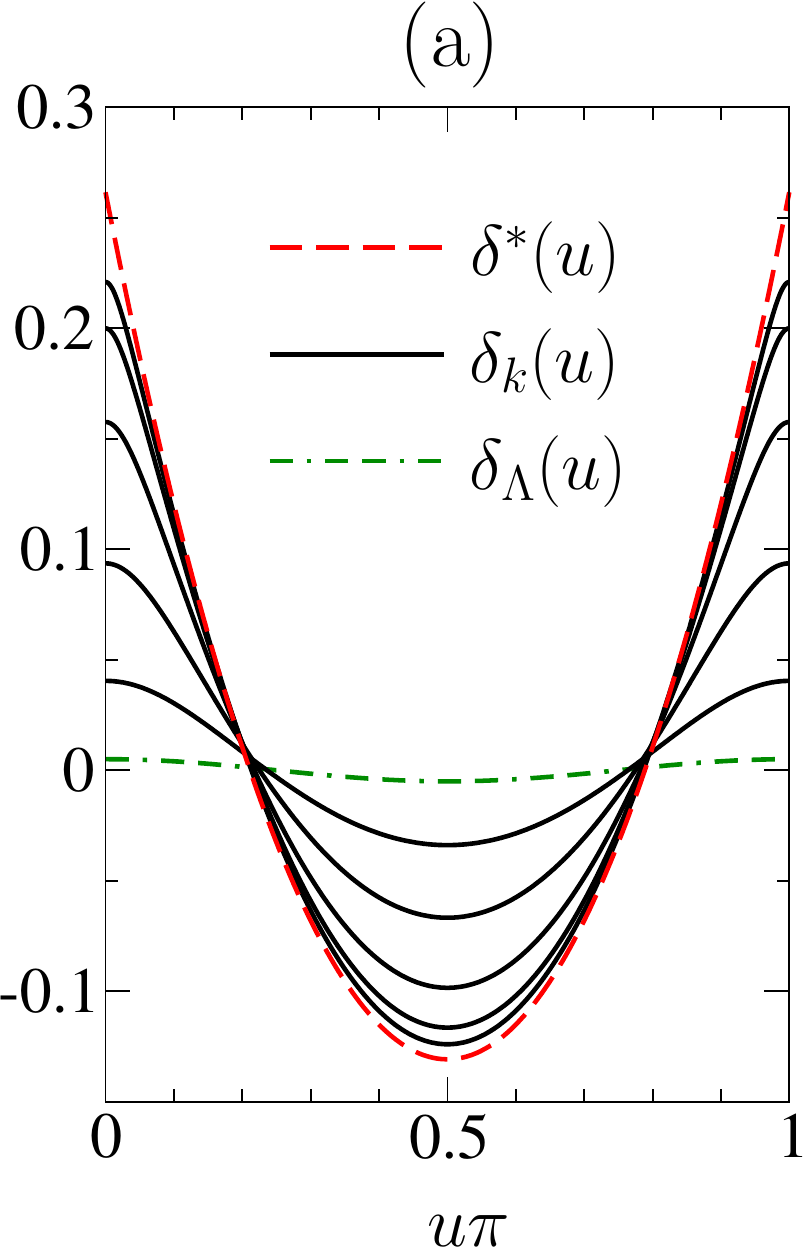}
			\hspace{0.25cm}
			\includegraphics[width=3.85cm,angle=0]{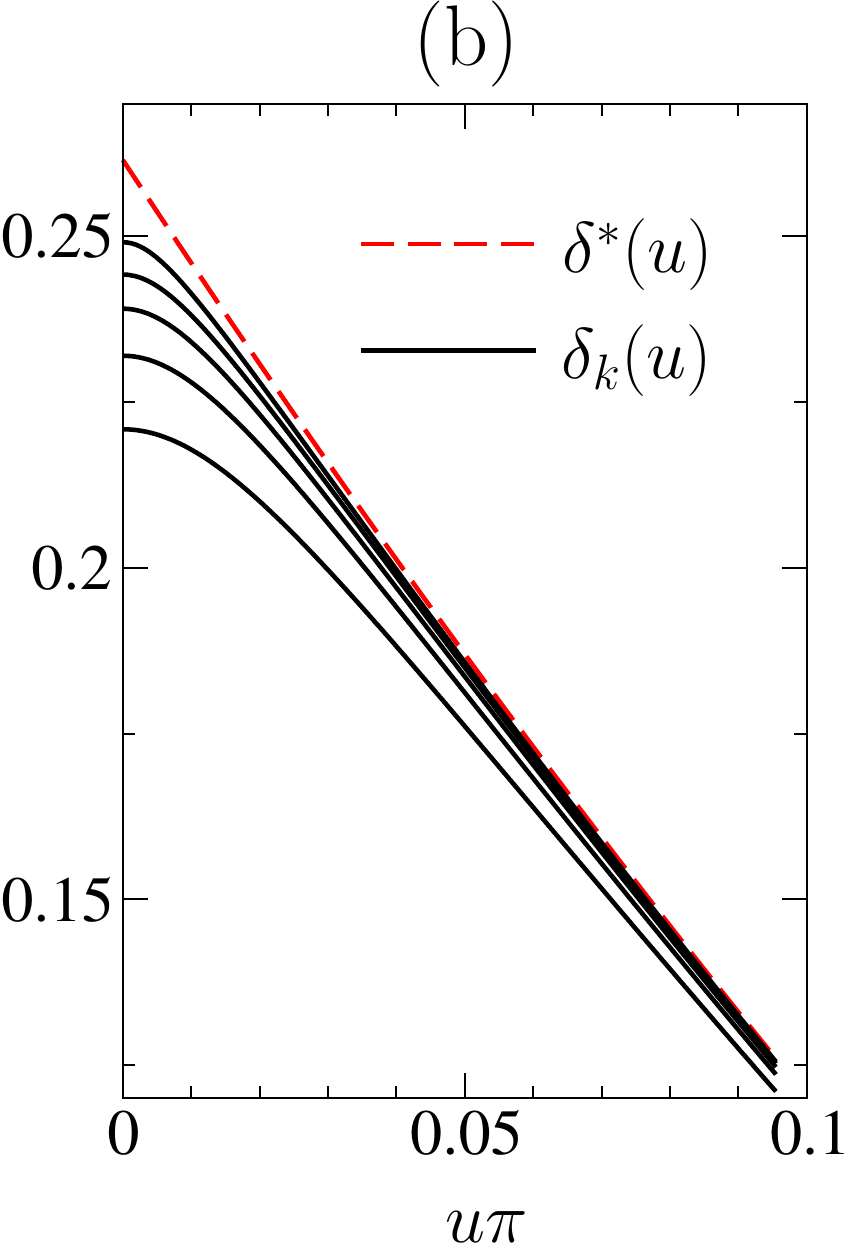}}
		\caption{(a) Potential $\delta_k(u)=-(K^2/v^2k^3)V_k''(u)$ for various values of $k$ ($K=0.4$ and $\delta_{1,\Lamb}=0.005$). The green curve shows the initial condition $\delta_\Lamb(u)=\delta_{1,\Lamb} \cos(2u)$ and the red one the fixed-point solution~(\ref{deltaFP}).
			(b) $\delta_k(u)$ for $u$ near 0 showing the formation of the quantum boundary layer ($k/\Lamb\simeq 0.050/0.030/0.018/0.011/0.007$ from bottom to top).} 
		\label{fig_pot} 
	\end{figure}
	
	For any nonzero momentum scale $k$, the cusp singularity is rounded in a quantum boundary layer as shown in Fig.~\ref{fig_pot}: for $u$ near 0, $\delta_k(0)-\delta_k(u)\propto |u|$ except in a boundary layer of size $|u|\sim K_k$; as a result the curvature $|\delta''_k(0)|\sim 1/K_k\sim k^{-\theta}$ diverges when $k\to 0$. 
	
	In the analogy with classical two-dimensional systems pointed out above, the fixed point describing the Bose-glass phase is a zero-temperature fixed point since the temperature $T_k\equiv K_k\sim k^\theta$ vanishes with $k$. In this context, the parabolic ``cuspy''  potential~(\ref{deltaFP}) and the boundary layer at nonzero scale $k$ have been obtained previously in the studies of random manifolds in disordered media or the random-field Ising model~\cite{Fisher86,Chauve00,Ledoussal02a,Ledoussal04,Ledoussal06a,Balents93,Balents04,Tarjus08,Tissier08,Tissier12,Tissier12a}. From a physical point of view, the cusp is due to the existence of metastable states leading to ``shock'' singularities (or static avalanches)~\cite{Balents96}: when the system is subjected to an external force, the ground state varies discontinuously whenever it becomes degenerate with a metastable state (which then becomes the new ground state)~\cite{[{At zero temperature the presence of the cusp implies the existence of a nonzero threshold force below which the system stays trapped in a minimum of the random potential and above which transitions between metastable states occur $via$ avalanches, see }]Chauve00}. At finite temperatures the system has a small but nonzero probability to be in two distinct, nearly degenerate, configurations (the rare excitations with energies of order of $T$ or smaller are thermally active), which results in a smearing of the cusp. 
	
	A similar interpretation holds in the Bose-glass phase. In the classical limit $K\to 0$ (corresponding to the $T\to 0$ limit of the classical model), the cusp in $\delta^*(u)$ is due to metastable states (defined as the minima of the action $S[\varphi]$ derived from $\hat H_0+\hat H_{\rm dis}$, i.e. before disorder averaging) becoming degenerate with the ground state. A nonzero value of $K$ leads to the possibility of quantum tunneling between different metastable configurations (a small number of low-energy metastable states become quantum-mechanically active) and a rounding of the cusp in a quantum boundary layer. These quantum tunneling events allow the system to escape pinning and one expects the existence of (rare) ``superfluid'' regions with significant density fluctuations and therefore reduced fluctuations (i.e., a nonzero rigidity) of the phase $\hat\theta$ of the boson operator $\hat\psi=e^{i\hat\theta}\sqrt{\hat\rho}$. (We further elaborate on that point below.)
	
\begin{figure}
		\centerline{\includegraphics[width=6.5cm]{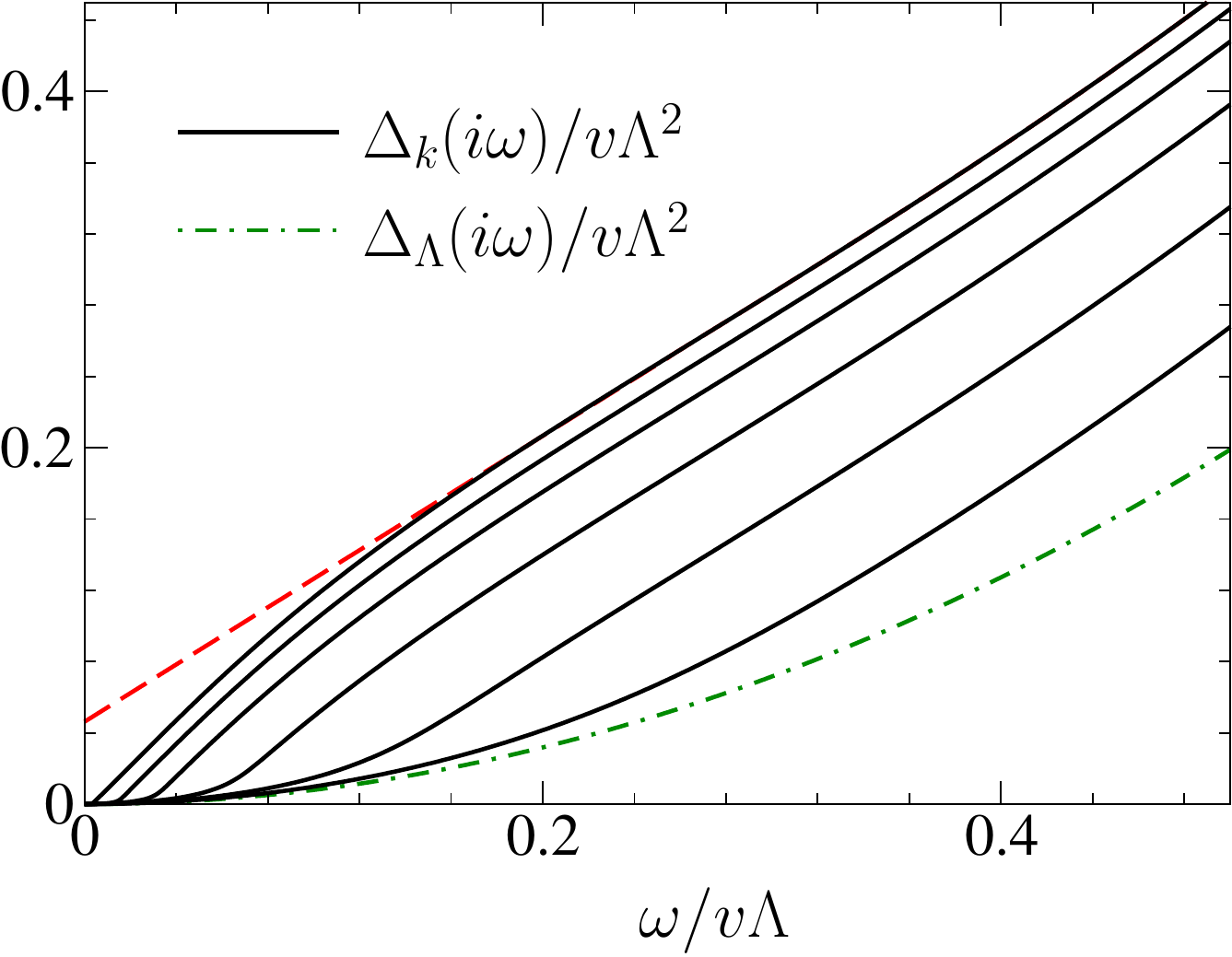}} 
		\caption{Low-frequency behavior of the self-energy $\Delta_k(i\w)$ for $k/\Lamb\simeq 1/0.223/0.135/0.082/0.050/0.030/0.011$
			(from bottom to top) in the case where $z=1+\theta=3/2$ ($K=0.4$ and $\delta_{1,\Lamb}=0.005$). The green dash-dotted line shows the initial condition $\Delta_\Lamb(i\w)=\w^2/\pi v K$ and the red dashed line is a guide to the eyes.} 
		\label{fig_Delta} 
	\end{figure}

	The thermal boundary layer of the two-dimensional classical model is associated with the existence of rare thermal excitations in the statics and activation barriers in the dynamics~\cite{Balents04}. Not surprisingly, we find that the quantum boundary layer controls the (quantum) dynamics of the $\varphi$ field in the boson problem~\cite{not77}. This is readily seen by the fact that $\dk\Delta_k(i\w)$ is proportional to $\delta_k''(u=0)$~\cite{not11}. Results of the numerical integration of the flow equations are shown in Fig.~\ref{fig_Delta}. For $|\w|\ll v_kk$ $\Delta_k(i\w)$ varies quadratically with $\w$: $\Delta_k(i\w)=Z_x\w^2/v_k^2$ with $v_k=\pi Z_xK_k\sim k^{\theta}$ when $k\to 0$~\cite{not100}. In the opposite limit $|\w|\gg v_kk$, when $k\to 0$ it is possible to find an analytical solution, $\Delta_k(i\w) = A + B |\w|^{(2-\theta)/z}$, where the positive constants $A$ and $B$ depend on the initial conditions of the flow at scale $k=\Lamb$ (see the red dashed line in Fig.~\ref{fig_Delta}). We therefore conclude that the self-energy converges nonuniformly towards a singular solution~\cite{not99}: 
	\begin{equation}
	\lim_{k\to 0} \Delta_k(i\w) = \llbrace 
	\begin{array}{lll} 
	0 & \mbox{if} & \w=0 , 
	\\ A+B|\w|^{(2-\theta)/z} & \mbox{if} & \w\neq 0 ,
	\end{array}
	\right. 
	\label{Delta}
	\end{equation} 
	in the low-energy limit. 
	
	The conductivity is given by 
	\beq
	\sig(\w) = - \frac{i\w}{\pi^2\Delta(\w)} 
	= \xi^2 \kappa (-i\w+\w^2\tau_c) +\calO(\w^3) ,
	\eeq 
	where $\Delta(\w)\equiv\Delta_{k=0}(i\w\to\w+i0^+)$ is the retarded self-energy, $\xi=\sqrt{Z_x/A}$ the pinning (or localization) length and $\tau_c=B/A$ the associated time scale. In the classical limit $K\to 0$ the compressibility $\kappa=K/\pi v$ vanishes and the conductivity is $\calO(\w^3)$. At nonzero $K$, our calculation shows that the low-frequency transport in the Bose-glass phase is due to the quantum tunneling events between different metastable configurations (and thus the rare superfluid regions) encoded in the quantum boundary layer.
	
	A phenomenological description of glassy classical systems is provided by the droplet scenario~\cite{Fisher88b,Balents04,Balents05}. The latter supposes the existence, at each length scale $L$, of a small number of excitations above the ground state, drawn from an energy distribution of width $\Delta E\sim L^\theta$ with a constant weight $\sim L^{-\theta}$ near $E=0$. The number of thermally active excitations is therefore $\sim T/L^\theta$, i.e., the system has a probability $\sim T/L^\theta$ to be in two nearly degenerate configurations. Thermal fluctuations are dominated by these rare droplet excitations and one has 
	$\overline{ [\mean{\varphi^2}-\mean{\varphi}^2]^p} \sim (T/L^\theta)\ln L$ 
	at length scale $L$~\cite{not110}. We have verified that this relation holds for $p=1$ and $p=2$ in the Bose-glass phase (with $T\equiv K$) thus validating the droplet picture. Although our approach is justified only in the limit of weak disorder~\cite{Giamarchi87,*Giamarchi88}, the low-energy physics of the Bose-glass phase is is expected to be independent of the disorder strength~\cite{Pollet14} so that the droplet scenario should hold in the entire localized phase~\cite{not130}. 
	
	Let us finally justify the choice $z=1+\theta=3/2$. The FRG approach yields a value of $\theta$ which is universal, i.e. independent of the microscopic parameters of the model, but strongly dependent on the regulator function $R_k(q,i\w)$ used in the implementation of the RG approach for reasons discussed in the Supplemental Materials~\cite{not11}. For a generic value of $\theta$ one would find $\Re[\sig(\w)] \sim |\w|^{3/z}$. By choosing $\theta=1/2$, which is achieved with a fine tuning of $R_k$, one ensures that the exact result $\Re[\sig(\w)]\sim \w^2\ln^2|\w|$ when $K=1$~\cite{Giamarchi_book} (corresponding to hard-core bosons or free fermions) is reproduced up to logarithmic corrections.

	\paragraph{Conclusion.}
	
	We have shown that the FRG description of classical disordered systems extends to the Bose-glass phase of one-dimensional Bose fluids. A necessary condition, however, is to use a nonperturbative approach in order to be able to reach the strong-disorder RG fixed point which characterizes the Bose-glass phase. Many of our results, in particular for the statics, are similar to those obtained in classical disordered systems (in which the long-distance physics is controlled by a zero-temperature fixed point). A key feature is the cuspy functional form of the disorder correlator which reveals the existence of metastable states and the ensuing glassy properties such as the presence of ``shocks'' (static avalanches). The presence of a quantum boundary layer rounding the cusp at nonzero momentum scale $k$, and encoding the quantum tunneling events between different metastable states (i.e. the rare superfluid regions), is responsible for the $\w^2$ dependence of the conductivity at low frequencies. These results agree with the phenomenological droplet picture of glassy systems~\cite{Fisher88b}. It remains to be seen whether the droplet picture also holds in higher dimensions or is specific to the one-dimensional case.
	
	Numerical studies of disordered systems where the physics is dominated by rare regions are notoriously difficult. Yet shocks have been observed in numerical simulations of disordered systems~\cite{Middleton07,Ledoussal09a} and glassy properties have been recently numerically demonstrated in a two-dimensional fermionic Anderson insulator~\cite{Lemarie19}. We expect similar numerical studies to be possible in the Bose-glass phase of a one-dimension Bose fluid.

	\paragraph{Acknowledgments.}
	
	\begin{acknowledgments}
		I am indebted to P. Azaria, F. Cr\'epin, N. Laflorencie, G. Lemari\'e, E. Orignac, A. Ran\c{c}on, B. Svistunov, G. Tarjus, M. Tissier and K. Wiese for insightful discussions and/or a critical reading of the manuscript.   
	\end{acknowledgments} 
	
	\vspace{-0.5cm}
	
	%

\setcounter{equation}{0}
\newpage

\def\kin{k_{\rm in}} 
\def\Msol{M_{\rm sol}} 
\def\Mcl{M_{\rm cl}} 

\begin{center}
{\bf \Large Supplemental Material}
\end{center}

A one-dimensional disordered Bose fluid is described by the replicated Euclidean action 
\begin{multline}
S = \sum_a \int dx \inttau \frac{v}{2\pi K} \left\{ (\dx\varphi_a)^2 + \frac{(\dtau\varphi_a)^2}{v^2} \right\} \\ 
-\calD \sum_{a,b} \int dx \inttau \, d\tau' \cos[2\varphi_a(x,\tau)-2\varphi_b(x,\tau')] , 
\label{action1} 
\end{multline}
where a UV momentum cutoff is implied on momentum and frequency ($|q|,|\w|/v\leq\Lamb$). $a,b=1\cdots n$ are replica indices. The first term in~(\ref{action1}) corresponds to the clean system with $v$ the sound-mode velocity and $K$ the Luttinger parameter. The phase field $\varphi_a$ is related to the density by $\rho_a=\rho_0 - \frac{1}{\pi}\dx\varphi_a+2\rho_2 \cos(2\pi\rho_0 x-2\varphi_a)$ ($\rho_0$ is the average density).
The second term in~(\ref{action1}) comes from disorder averaging with $\calD=\rho_2^2D_b$ and $D_b$ the variance of the Gaussian random potential which couples to the density.

\section{Exact flow equation} 

To implement the FRG approach we add to the action the infrared regulator term 
\begin{equation}
\Delta S_k[\varphi] = \half \sum_{q,\w,a} \varphi_a(-q,-i\w) R_k(q,i\w) \varphi_a(q,i\w) 
\label{DeltaS}
\end{equation}
such that fluctuations are smoothly taken into account as $k$ is lowered from the microscopic scale $\Lamb$ down to 0~\cite{Berges02,Delamotte12,Kopietz_book}. The regulator function in~(\ref{DeltaS}) is defined by 
\begin{equation}
R_k(q,i\w) = ( Z_x q^2 + \Delta_k(i\w)) r\left( \frac{Z_x q^2 + \Delta_k(i\w)}{Z_xk^2} \right) ,
\label{Rk}  
\end{equation}
where $r(x)=\alpha/(e^x-1)$ with $\alpha$ a parameter of order unity. $Z_x$ and $\Delta_k(i\w)$ are defined below. Thus $\Delta S_k[\varphi]$ suppresses fluctuations such that $q^2\ll k^2$ and $\Delta_k(i\w)\ll Z_xk^2$ but leaves unaffected those with $q^2\gg k^2$ or $\Delta_k(i\w)\gg Z_xk^2$. 

The partition function of the replicated system, 
\begin{equation}
\calZ_k[J] = \int \calD[\varphi]\, e^{-S[\varphi]-\Delta S_k[\varphi] + \sum_a \inttau \int dx  J_a \varphi_a } , 
\end{equation}
is $k$ dependent. The scale-dependent effective action 
\begin{equation}
\Gamma_k[\phi] = -\ln\calZ_k[J] + \sum_a \inttau \int dx \, J_a \phi_a - \Delta S_k[\phi]
\end{equation}
is defined as a modified Legendre transform of $\ln\calZ_k[J]$ which includes the subtraction of $\Delta S_k[\phi]$. Here $\phi_a(x,\tau)=\mean{\varphi_a(x,\tau)}=\delta\ln\calZ_k[J]/\delta J_a(x,\tau)$ is the expectation value of the phase field (in the presence of the external source $J_a$). Assuming that all fluctuations are frozen by $\Delta S_\Lamb[\varphi]$, $\Gamma_\Lamb[\phi]=S[\phi]$ (as in mean-field theory). On the other hand the effective action of the original model is given by $\Gamma_{k=0}$ since $\Delta S_{k=0}=0$. The nonperturbative FRG approach aims at determining $\Gamma_{k=0}$ from $\Gamma_\Lamb$ using Wetterich's equation~\cite{Wetterich93} 
\begin{equation}
\dt \Gamma_k[\phi] = \half \Tr \Bigl\{ \dt R_k \bigl( \Gamma_k^{(2)}[\phi] + R_k )^{-1} \Bigr\} , 
\label{eqwet} 
\end{equation}
where $t=\ln(k/\Lamb)$ is a (negative) RG ``time''.

\section{Free replica sum expansion} 

A common approximation to solve the flow equation~(\ref{eqwet}) is to expand the effective action 
\begin{equation}
\Gamma_k[\phi] = \sum_a \Gamma_{1,k}[\phi_a] - \half \sum_{a,b} \Gamma_{2,k}[\phi_a,\phi_b] + \cdots 
\end{equation}
in increasing number of free replica sums. In the following we retain only $\Gamma_{1,k}$ and $\Gamma_{2,k}$ and consider the ansatz 
\begin{equation}
\begin{split} 
&\Gamma_{1,k}[\phi_a] = \int dx \inttau \llbrace \frac{Z_{x}}{2} (\dx\phi_a)^2 + \half \phi_a \Delta_k(-\dtau) \phi_a \rrbrace , \\ 
&\Gamma_{2,k}[\phi_a,\phi_b] = \int dx \inttau\, d\tau'\, V_k(\phi_a(x,\tau)-\phi_b(x,\tau')) , 
\end{split}
\label{ansatz1}
\end{equation}
with initial conditions $\Delta_\Lamb(i\w)=\w^2/\pi vK$ and $V_\Lamb(u)=2\calD \cos(2u)$. $V_k(u)$ is a $\pi$-periodic function. The form of $\Gamma_{1,k}$ is strongly constrained by the statistical tilt symmetry, i.e. the invariance of the disorder part of the action~(\ref{action1}) in the change $\varphi_a(x,\tau)\to\varphi_a(x,\tau)+w(x)$: $Z_x$ remains equal to its initial value $v/\pi K$ and no higher-order space derivatives are allowed, the self-energy $\Delta_k(i\w)$ must satisfy $\Delta_k(i\w=0)=0$ and the two-replica potential $V_k$ is a function of $\phi_a(x,\tau)-\phi_b(x,\tau')$. By inserting the ansatz~(\ref{ansatz1}) into~(\ref{eqwet}) we obtain coupled RG equations for $\Delta_k(i\w)$ and $V_k(u)$. 

In practice we consider the dimensionless functions 
\begin{equation}
\delta_k(u) = - \frac{K^2}{v^2} \frac{V_k''(u)}{k^3} , \quad 
\tilde \Delta_k(i\tw) = \frac{\Delta_k(i\w)}{Z_xk^2} , 
\end{equation} 
where $\tw=\w/v_kk$ is a dimensionless frequency and $v_k$ the running velocity obtained from the low-energy behavior of the self-energy: $\Delta_k(i\w)=Z_x\w^2/v_k^2+\calO(\w^4)$. It is also convenient to define a running Luttinger parameter $K_k$ {\it via} $Z_x=v_k/\pi K_k$. The flow equations then read 
\begin{equation}
\begin{split} 
\dt\delta_k(u) ={}& -3 \delta_k(u) - K_k l_1 \delta''_k(u) \\ & 
+ \pi \bar l_2 [ \delta_k''(u) (\delta_k(u)-\delta_k(0)) + \delta'_k(u)^2 ] , \\ 
\dt \tilde\Delta_k(i\tw) ={}& - 2 \tilde\Delta_k(i\tw) + z_k \tw \partial_{\tw} \tilde\Delta_k(i\tw) \\ &
- \pi \delta''_k(0) [ \bar l_1(\tw) - \bar l_1(0) ] , \\ 
\dt K_k ={}& \theta_k K_k , \qquad
\dt (K_k/v_k) = 0 , 
\end{split}
\label{rgeq} 
\end{equation}
where $z_k=1+\theta_k$ is the running dynamical critical exponent and  
\begin{equation}
\theta_k = \frac{\pi}{2} \delta''_k(0) \bar m_\tau . 
\label{thetak} 
\end{equation}
The ``threshold'' functions $l_1,\bar l_2,\bar l_1(\tw)$  are defined in Sec.~\ref{sec_threshold}. $\bar l_2$ and $\bar l_1(0)$ are $k$ independent while $l_1$ and $\bar l_1(\tw\neq 0)$ depend on $\theta_k$.

\subsection*{Weak-disorder limit} 

In the weak-disorder limit it is sufficient to approximate $\Delta_k(i\w)=Z_x\w^2/v_k^2$ and $V_k(u)=2\calD_k \cos(2u)$, i.e. $\delta_k(u)=\delta_{1,k}\cos(2u)$ with $\delta_{1,k}=8K^2\calD_k/v^2k^3$. This gives 
\beq
\begin{split}
	\dt \delta_{1,k} &= (-3+4K_kl_1)\delta_{1,k} + 4\pi \bar l_2 \delta_{1,k}^2 ,  \\
	\theta_k &= -2\pi \bar m_\tau \delta_{1,k} ,
\end{split} 
\label{flow1} 
\eeq
where $\bar m_\tau= \partial_{\tw^2} \bar l_1(\tw) \bigl|_{\tw=0}$. Since $l_1=1/2$ for $\theta_k=0$, $K_k=K$, $v_k=v$ and $\Delta_k(i\w)=\w^2/\pi vK$ (i.e. $\tilde\Delta_k(i\tw)=\tw^2$), one sees that the superfluid phase is stable against (infinitesimal) disorder when $K>3/2$. In the vicinity of $(K_k=3/2,\delta_{1,k}=0)$, to leading order Eqs.~(\ref{flow1}) become  
\beq
\begin{split}
	\dt \delta_{1,k} &= 2\delta_{1,k}(K_k-3/2) ,  \\ 
	\dt K_k &= -3\pi \bar m_\tau \delta_{1,k},
\end{split} 
\eeq
where $\bar m_\tau\equiv \bar m_\tau|_{\theta_k=0}<0$. These equations are similar to those obtained in Refs.~\onlinecite{Giamarchi87,*Giamarchi88,Ristivojevic12,*Ristivojevic14}.

\section{Bose-glass phase} 

\subsection{A. Fixed-point potential} 

In addition to the line of trivial fixed points $\delta^*(u)=0$, which is stable only for $K^*>3/2$ and corresponds to the superfluid phase, the flow equations of $K_k$ and $\delta_k(u)$ admit a nontrivial fixed point defined by $K^*=0$ and 
\begin{equation}
\delta^*(u) = \frac{1}{2\pi\bar l_2} \left[ \left( u - \frac{\pi}{2} \right)^2 - \frac{\pi^2}{12} \right] \quad \mbox{for} \quad u\in [0,\pi] . 
\label{deltafp} 
\end{equation}
Thus $\delta^*(u)$ exhibits a cusp at $u=n\pi$ ($n\in\mathbb{Z}$). A stability analysis shows that this fixed point is fully attractive. 

For any nonzero momentum scale $k$, the cusp singularity is rounded in a quantum boundary layer: for $u$ near 0, $\delta_k(0)-\delta_k(u)\propto |u|$ except in a boundary layer of size $|u|\sim K_k$; as a result the curvature $|\delta''_k(0)|\sim 1/K_k\sim k^{-\theta}$, diverges when $k\to 0$ with the exponent $\theta=\lim_{k\to 0}\theta_k$.

\subsection{B. Self-energy}

By definition of the dynamical critical exponent $z_k$, $\tilde\Delta_k(i\tw)=\tw^2+\calO(\tw^4)$ for $|\tw|\ll 1$. In the opposite limit $|\tw|\gg 1$, the threshold function $\bar l_1(i\tw)$ can be neglected in the equation $\dt\tilde\Delta_k(i\tw)$ and we obtain 
\begin{equation}
\dt \tilde\Delta_k(i\tw) 
=  - 2 \tilde\Delta_k(i\tw) + z \tw \partial_{\tw} \tilde\Delta_k(i\tw) 
- \frac{\pi C}{K_k} \bar l_1(0) ,
\end{equation}
where, assuming $k$ small enough, we have approximated $\delta_k''(0)$ by $-C/K_k$ (with $C$ a positive constant) and $z_k$ by its fixed-point value $z=\lim_{k\to 0}z_k=1+\theta$. Looking for a solution in the form 
\begin{equation}
\tilde\Delta_k(i\tw) = \tilde A_k + \frac{\tilde B}{K_k} |\tw|^x , 
\end{equation}
we find
\begin{equation}
x = \frac{2-\theta}{z} , \quad \tilde A_k = \frac{\tilde A}{k^2} - \frac{1}{K_k} \frac{\pi C\bar l_1}{2-\theta} , 
\end{equation} 
where $\tilde A$ and $\tilde B$ are $k$ independent. 
This implies that for $k\to 0$ and $|\w|\gg v_kk$ the self-energy $\Delta_k(i\w)$ takes the form 
\begin{equation}
\Delta_k(i\w) = Z_x k^2 \frac{\tilde B}{K_k} \left( \frac{|\w|}{\pi Z_x k K_k} \right)^\frac{2-\theta}{z} + Z_x \tilde A - \frac{Z_xk^2}{K_k} \frac{\pi C\bar l_1}{2-\theta}  
\end{equation}
in the low-frequency limit. 
Since $K_k\sim k^\theta$ with $\theta<2$ the last term in the rhs tends to zero for $k\to 0$ while the first one becomes $k$ independent, i.e. 
\begin{equation}
\Delta_k(i\w) = A + B |\w|^\frac{2-\theta}{z} .
\end{equation} 

\subsection{C. The exponent $\theta$} 

Since the threshold function $\bar m_\tau\equiv\bar m_\tau(\theta_k)$ is a linear function of $\theta_k$ (Sec.~IV), Eq.~(\ref{thetak}) gives 
\begin{equation} 
\theta_k = \frac{\pi}{2} \frac{\delta_k''(0) \bar m_\tau(0)}{1-\frac{\pi}{2} \delta_k''(0) [\bar m_\tau(1) - \bar m_\tau(0)]} 
\end{equation}
and 
\begin{equation}
\theta = \lim_{k\to 0} \theta_k = \frac{\bar m_\tau(0)}{\bar m_\tau(0) - \bar m_\tau(1)}
\label{theta}
\end{equation}
using $\lim_{k\to 0}\delta_k''(0)=-\infty$. With the regulator function~(\ref{Rk}) and $r(x)=\alpha/(1-e^x)$ one finds that $\theta$ decreases from 0.76 to 0.26 when $\alpha$ increases from 2 to 3; there is no principle of minimum sensitivity which would allow one to determine the optimal choice of $\alpha$. The value $\theta=1/2$ is obtained for $\alpha\simeq 2.422$.

This strong dependence on the regulator function is an unavoidable consequence of~(\ref{theta}) and is in sharp contrast with usual second-order phase transitions where the critical exponents depend on both the threshold functions and the values of the coupling constants at the fixed point. In the latter case one observes that the dependence of the coupling constants on $R_k$ largely compensates that of the threshold functions to make the critical exponents eventually weakly dependent on the regulator function.

\section{Threshold functions}
\label{sec_threshold}

The threshold functions are defined by 
\begin{equation}
\begin{gathered}
l_n = n \int_0^\infty d\tilde q \int_{-\infty}^\infty \frac{d\tw}{2\pi} \frac{\dt R_k(\tilde q,i\tw)}{Z_xk^2} 
\tilde G(\tilde q,\tw)^{n+1} , \\ 
\bar l_n(i\tw) = n \int_0^\infty d\tilde q \frac{\dt R_k(\tilde q,i\tw)}{Z_xk^2} 
\tilde G(\tilde q,\tw)^{n+1} , \\  
\bar m_\tau = \partial_{\tw^2} \bar l_1(\tw) \bigl|_{\tw=0} ,
\end{gathered}
\end{equation}
where 
\begin{align} 
\tilde G(\tilde q,\tw) ={}& \frac{1}{(\tilde q^2+\tilde\Delta_k)(1+r)} , \nonumber \\  
\frac{\dt R_k(\tilde q,i\tw)}{Z_xk^2}  ={}& 2 \tilde\Delta_k r- 2 (\tilde q^2+\tilde\Delta_k) \tilde q^2 r' \nonumber \\ & \hspace{-1cm}
+ (\dt\tilde\Delta_k|_\tw - z_k \tw \partial_\tw \tilde\Delta_k)[r + (\tilde q^2+\tilde\Delta_k)r'] ,
\end{align}
with $r\equiv r(\tilde q^2+\tilde\Delta_k)$ and $\tilde\Delta_k\equiv\tilde\Delta_k(i\tw)$. For $\theta_k=0$ and $\tilde\Delta_k(i\tw)=\tw^2$, the threshold function $l_1=1/2$ is universal, i.e. independent of the function $r(x)$ provided that the latter satisfies $r(0)=\infty$ and $r(\infty)=0$.


\end{document}

%% file: definition.tex

\def\rhoeq{\hat\rho_{\rm eq}}

\newcommand{\marge}[1]{\marginpar{\scriptsize #1}}
\newcommand{\remarque}[1]{\marginpar{\scriptsize Remarque}{\it [#1]}}
\newcommand{\new}[1]{{\bf #1}}
\newlength{\textlarg}
\newcommand{\redbar}[1]{\textcolor{red}{\st{#1}}} 
\newcommand{\bluebar}[1]{\textcolor{blue}{\st{#1}}} 

\newcommand{\beq}{\begin{equation}}
\newcommand{\eeq}{\end{equation}}
\newcommand{\bleq}{\begin{eqnarray}}
\newcommand{\eleq}{\end{eqnarray}} 
\newcommand{\bfig}{\begin{figure}}
\newcommand{\efig}{\end{figure}}
\newcommand{\bline}{\begin{multline}}
\newcommand{\eline}{\end{multline}}
\newcommand{\bremark}{\begin{quotation} \noindent \small }
\newcommand{\eremark}{\end{quotation}}
\newcommand{\llbrace}{\left\lbrace}  
\newcommand{\rrbrace}{\right\rbrace}
\newcommand{\lbraket}{\left[}
\newcommand{\rbraket}{\right]}
\newcommand{\llangle}{\left\langle}
\newcommand{\rrangle}{\right\rangle} 

\newcommand{\Tr}{{\rm Tr}} 
\newcommand{\tr}{{\rm tr}} 
\newcommand{\sgn}{\,{\rm sgn}} 
\newcommand{\mean}[1]{\langle #1 \rangle}
\newcommand{\commu}[2]{[#1,#2]} 
\newcommand{\bra}[1]{\langle#1|}
\newcommand{\ket}[1]{|#1\rangle}
\newcommand{\braket}[2]{\langle #1|#2\rangle}
\newcommand{\ketbra}[2]{|#1\rangle\langle#2|}
\newcommand{\dbraket}[3]{\langle #1|#2|#3\rangle}
\newcommand{\tens}[1]{\overleftrightarrow{#1}}  
\newcommand{\vac}{|{\rm vac}\rangle} 
\newcommand{\bravac}{\langle{\rm vac}|}
\newcommand{\const}{{\rm const}} 
\newcommand{\atanh}{\,{\rm atanh}}
\newcommand{\cotanh}{\,{\rm cotanh}}

\newcommand{\ie}{i.e.\xspace}
\newcommand{\iet}{i.e.}
\newcommand{\eg}{e.g.\xspace}
\newcommand{\cc}{{\rm c.c.}} 
\newcommand{\hc}{{\rm h.c.}} 
\newcommand{\etal}{{\it et al. }}
\newcommand\eme{$^{\mbox{\footnotesize ème}}$\xspace}

\newcommand{\jhatbf}{\hat {\textbf \jold}} 
\newcommand{\Jhatbf}{\hat {\textbf \J}} 
\newcommand{\jhat}{\hat {\jmath}} 
\newcommand{\Jhat}{\hat {J}} 
\newcommand{\jbf}{\textbf j}
\newcommand{\Jbf}{\textbf J}

\def\chibf{\boldsymbol{\chi}}
\def\down{\downarrow}
\def\eps{\epsilon}
\def\gam{\gamma} 
\def\alphabf{\boldsymbol{\alpha}}
\def\phibf{\boldsymbol{\phi}}
\def\varphibf{\boldsymbol{\varphi}}
\def\varphibfs{\boldsymbol{\varphi}_<}
\def\varphibfl{\boldsymbol{\varphi}_>}
\def\varphis{\varphi_{<}}
\def\varphil{\varphi_{>}}
\def\psibf{\boldsymbol{\psi}}
\def\thetabf{\boldsymbol{\theta}}
\def\Ome{\Omega}
\def\omeD{{\omega_D}} 
\def\bfOme{\boldsymbol{\Omega}} 
\def\Omebf{\boldsymbol{\Omega}} 
\def\lamb{\lambda}
\def\Lamb{\Lambda}
\def\sig{\sigma}
\def\Sig{\Sigma}
\def\sigp{{\sigma'}} 
\def\bfsig{\boldsymbol{\sigma}} 
\def\sigbf{\boldsymbol{\sigma}} 
\def\bfSig{\boldsymbol{\Sigma}} 
\def\The{\Theta} 
\def\up{\uparrow}

\def\epsk{\epsilon_{\bf k}} 
\def\xik{\xi_{\bf k}} 
\def\txik{\tilde\xi_{\bf k}} 
\def\xip{\xi_{\bf p}} 
\def\xiq{\xi_{\bf q}} 
\def\xikq{\xi_{{\bf k}+{\bf q}}} 
\def\Ek{E_{\bf k}} 
\def\Ep{E_{\bf p}}
\def\Eq{E_{\bf q}}
\def\Heff{\hat H_{\rm eff}}
\def\Hem{\hat H_{\rm em}}
\def\Hint{\hat H_{\rm int}}
\def\Hloc{\hat H_{\rm loc}}
\def\HMF{\hat H_{\rm MF}}
\def\Sem{S_{\rm em}}
\def\SMF{S_{\rm MF}} 
\def\SHF{S_{\rm HF}} 
\def\SRPA{S_{\rm RPA}} 
\def\Sint{S_{\rm int}} 
\def\Sloc{S_{\rm loc}}
\def\TN{T_{\rm N}} 
\def\TNHF{T^{\rm HF}_{\rm N}} 
\def\Zloc{Z_{\rm loc}} 
\def\ZMF{Z_{\rm MF}} 
\def\ZHF{Z_{\rm HF}} 
\def\ZRPA{Z_{\rm RPA}} 
\def\RPA{{\rm RPA}}
\def\loc{{\rm loc}} 
\def\pp{{\rm pp}}
\def\ph{{\rm ph}} 
\def\ch{{\rm ch}}
\def\sp{{\rm sp}} 
\def\qtf{q_{\rm TF}}
\def\epstf{\eps^{}_{\rm TF}} 
\def\epsrpa{\eps^{}_{\rm RPA}} 
\def\chinnzpp{\chi_{nn}^{0}{}\!\!\!''}

\def\half{\frac{1}{2}}
\def\dhalf{\dfrac{1}{2}}
\def\third{\frac{1}{3}} 
\def\quarter{\frac{1}{4}}

\def\qr{{\bf q}\cdot{\bf r}}
\def\wt{\omega t} 

\def\a{{\bf a}}
\def\b{{\bf b}}
\newcommand{\cv}{{\bf c}} 
\def\e{{\bf e}}
\def\f{{\bf f}}
\def\g{{\bf g}}
\def\h{{\bf h}}
\def\jold{\char"11}
\def\j{{\bf j}}
\def\k{{\bf k}}
\def\l{{\bf l}}
\def\m{{\bf m}}
\def\n{{\bf n}} 
\def\p{{\bf p}} 
\def\q{{\bf q}}
\def\r{{\bf r}}
\def\t{{\bf t}}
\def\u{{\bf u}}
\newcommand{\vv}{{\bf v}}
\def\x{{\bf x}}
\def\y{{\bf y}} 
\def\z{{\bf z}} 
\def\A{{\bf A}}
\def\B{{\bf B}}
\def\D{{\bf D}} 
\def\E{{\bf E}} 
\def\F{{\bf F}} 
\def\H{{\bf H}}  
\def\J{{\bf J}}
\def\K{{\bf K}} 

\def\G{{\bf G}}
\def\L{{\bf L}}
\def\M{{\bf M}}  
\def\O{{\bf O}} 
\def\P{{\bf P}} 
\def\Q{{\bf Q}} 
\def\R{{\bf R}}
\def\S{{\bf S}}
\def\U{{\bf U}} 
\def\V{{\bf V}} 
\def\X{{\bf X}} 
\def\Y{{\bf Y}} 
\def\epsbf{\boldsymbol{\epsilon}}
\def\betabf{\boldsymbol{\beta}}
\def\deltabf{\boldsymbol{\delta}}
\def\mubf{\boldsymbol{\mu}}
\def\nablabf{\boldsymbol{\nabla}}
\def\rhobf{\boldsymbol{\rho}}
\def\sigmabf{\boldsymbol{\sigma}} 
\def\Pibf{\boldsymbol{\Pi}}
\def\pibf{\boldsymbol{\pi}}

\def\para{\parallel}
\def\kpara{{k_\parallel}}
\def\kperp{{k_\perp}} 
\def\kperpp{{k_\perp'}} 
\def\qperp{{q_\perp}} 
\def\tperp{{t_\perp}} 

\def\w{\omega}
\def\wn{\omega_n}
\def\wm{\omega_m}
\def\wnu{\omega_\nu}
\def\wp{\omega_p} 
\def\dmu{{\partial_\mu}}
\def\dnu{{\partial_\nu}}
\def\dl{{\partial_l}}  
\def\dt{\partial_t} 
\def\tdt{\tilde\partial_t}
\def\dk{\partial_k}
\def\tdk{\tilde\partial_k}
\def\dx{\partial_x}
\def\dy{\partial_y} 
\def\dtau{{\partial_\tau}}  
\def\det{{\rm det}} 
\def\Pf{{\rm Pf}}
\def\diag{{\rm diag}}

\def\dsum{\displaystyle \sum}
\def\dint{\displaystyle \int} 
\def\intt{\int_{-\infty}^\infty dt} 
\def\inttp{\int_{-\infty}^\infty dt'} 
\def\intk{\int_{\bf k}} 
\def\intkd{\int \frac{d^dk}{(2\pi)^d}}
\def\intq{\int_{\bf q}} 
\def\intr{\int d^dr}  
\def\dintr{\displaystyle \int d^dr} 
\def\intrp{\int d^dr'}
\def\dinttau{\displaystyle \int_0^\beta d\tau}
\def\dinttaup{\displaystyle \int_0^\beta d\tau'}
\def\inttau{\int_0^\beta d\tau}
\def\inttaup{\int_0^\beta d\tau'}
\def\intx{\int d^{d+1}x} 
\def\inttaur{\int_0^\beta d\tau \int d^dr}
\def\intinf{\int_{-\infty}^\infty}
\def\dinttaur{\displaystyle \int_0^\beta d\tau \int d^dr}
\def\dintinf{\displaystyle \int_{-\infty}^\infty}
\def\intw{\int_{-\infty}^\infty \frac{d\w}{2\pi}}
\def\sumr{\sum_{\bf r}} 

\def\calA{{\cal A}}
\def\calAbf{\bm{{\cal A}}}
\def\calB{{\cal B}} 
\def\calC{{\cal C}} 
\def\dt{\partial_t}
\def\calD{{\cal D}}
\def\calE{{\cal E}}
\def\calF{{\cal F}} 
\def\calFbf{\bm{{\cal F}}}
\def\calG{{\cal G}}
\def\calH{{\cal H}}
\def\calI{{\cal I}}
\def\calJ{{\cal J}}
\def\calK{{\cal K}}
\def\calL{{\cal L}} 
\def\calM{{\cal M}} 
\def\calN{{\cal N}}
\def\calO{{\cal O}}
\def\calP{{\cal P}}  
\def\calR{{\cal R}} 
\def\calS{{\cal S}}
\def\calT{{\cal T}}
\def\calU{{\cal U}}
\def\calV{{\cal V}}
\def\calX{{\cal X}} 
\def\calY{{\cal Y}} 
\def\calZ{{\cal Z}} 

\def\calbfB{{\bf \cal B}}
\def\calbfF{{\bf \cal F}}

\def\tT{{\tilde T}}
\def\talpha{{\tilde\alpha}}
\def\tdelta{{\tilde\delta}}
\def\tDelta{{\tilde\Delta}}
\def\teta{{\tilde\eta}} 
\def\tlamb{{\tilde\lambda}}
\def\tmu{{\tilde\mu}}
\def\tphibf{{\tilde\phibf}}
\def\trho{{\tilde\rho}}
\def\tvarphibf{{\tilde\varphibf}} 
\def\tw{{\tilde\omega}}
\def\twn{{\tilde\omega_n}}
\def\twnu{{\tilde\omega_\nu}}

\def\asinh{{\rm asinh}} 

%% file: bg19.bbl
\begin{thebibliography}{66}
\bibitem [{not({\natexlab{a}})}]{not33}%
		\BibitemOpen
		\href@noop {} {} \bibinfo {note} {Our approach closely
			follows the nonperturbative FRG study of the random-field Ising model
			reported in Refs.~\cite{Tarjus08,Tissier08}.}\BibitemShut {Stop}%
\bibitem [{not({\natexlab{d}})}]{not77}%
		\BibitemOpen
		\href@noop {} {} \bibinfo {note} {Note that contrary to
			the classical model the dynamics of the Bose fluid is entirely determined by
			the quantum Hamiltonian and there is no need to consider an additional
			equation (e.g. a Langevin equation).}\BibitemShut {Stop}%
\bibitem [{not({\natexlab{e}})}]{not100}%
		\BibitemOpen
		\href@noop {} {} \bibinfo {note} {This is a direct
			consequence of the infrared regulator used in the FRG implementation, which
			ensures that the self-energy $\Delta_k(i\w)$ is a regular function of $\w$
			near $\w=0$ for any nonzero $k$.}\BibitemShut {Stop}%
\bibitem [{not({\natexlab{f}})}]{not99}%
		\BibitemOpen
		\href@noop {} {} \bibinfo {note} {This singular
			self-energy was also found in the Gaussian variational method where the
			Bose-glass phase is described by a replica-symmetry-broken solution confined
			to the $\w=0$ mode~\cite{Giamarchi96}.}\BibitemShut {Stop}%
\bibitem [{not({\natexlab{g}})}]{not110}%
		\BibitemOpen
		\href@noop {} {} \bibinfo {note} {This equation is valid
			for systems with a vanishing roughness exponent $\zeta=0$. In the Bose-glass
			phase, this condition follows from the field $\varphi$ having a vanishing
			scaling dimension.}\BibitemShut {Stop}%
\bibitem [{not({\natexlab{h}})}]{not130}%
		\BibitemOpen
		\href@noop {} {} \bibinfo {note} {Note that the droplet
			picture, although reminiscent of, differs from the description of the
			Bose-glass phase as a Griffiths phase, as often advocated in particular near
			the transition to the Mott insulating
			phase~\cite{Fisher89,Pollet09,Niederle13,Hegg13}. In the Griffiths phase the
			superfluid regions are exponentially rare while they are power-law rare in
			the droplet scenario.}\BibitemShut {Stop}%
\end{thebibliography}
